\documentclass[aps,prb,showpacs,twocolumn,amsmath,amssymb,superscriptaddress]{revtex4}
\usepackage{epsfig}
\usepackage{bm}
\def\be{\begin{equation}}
\def\ee{\end{equation}}
\def\bea{\begin{eqnarray}}
\def\eea{\end{eqnarray}}
\def\la{\langle}
\def\ra{\rangle}
\begin{document}
\title{Qubit feedback and control with kicked quantum nondemolition
measurements:\\ A quantum Bayesian analysis}
\author{Andrew N. Jordan}
\affiliation{D\'epartement de Physique Th\'eorique, Universit\'e
de Gen\`eve, CH-1211 Gen\`eve 4, Switzerland}
\affiliation{Institute for Quantum Studies, Texas A\&M University,
College Station, TX 77843-4242, USA}
\author{Alexander N. Korotkov}
\affiliation{ Department of Electrical Engineering, University of
California, Riverside, CA 92521-0204, USA}
\date{\today}
\begin{abstract}
The informational approach to continuous quantum measurement is
derived from POVM formalism for a mesoscopic scattering detector
measuring a charge qubit. Quantum Bayesian equations for the qubit
density matrix are derived, and cast into the form of a stochastic
conformal map. Measurement statistics are derived for kicked
quantum nondemolition measurements, combined with conditional
unitary operations. These results are applied to derive a feedback
protocol to produce an arbitrary pure state after a weak
measurement, as well as to investigate how an initially mixed
state becomes purified with and without feedback.
\end{abstract}
 \maketitle

\section{Introduction}
Quantum measurement is usually taught in textbooks as an
instantaneous process.  However, in nature, all processes take a
finite time.  Physical projective measurement (or wave-function
collapse) must therefore happen over some time period, and is
often a sequence of {\it weak measurements}, run for a
sufficiently long time. Weak quantum measurements are
characterized by an intrinsic uncertainty about the state of the
measured system.  In the parlance of detector physics, this is
equivalent to the statement that the signal cannot be confidently
distinguished from the noise without a sufficiently long
integration time.  At some intermediate time, the eigenstates of
the measurement operator can only be assigned a value with some
confidence, determined with the probability of a given
realization of the detector output (similarly to classical
Bayesian inference).

In solid state systems, the typically weak coupling constant
between the quantum system and measuring apparatus imply that weak
measurements with long measurement times are the norm.  While this
is often frustrating to experimentalists who wish to preform
projective measurement for quantum computation purposes, we view
this situation as an opportunity to discover and implement ideas
in quantum measurement that are qualitatively different from
standard projective measurement. Although many results in this
paper are abstract and apply to many different physical systems
and detectors, the results will be discussed in terms of solid
state physics.  Quantum detection in mesoscopic
structures began with the ``controlled dephasing" experiments of 
Ref.~\onlinecite{dephaseexp} and related theoretical works,\cite{dephaseth} 
which has continued to be an active area
of research.\cite{thother}  The particular mesoscopic structure we shall focus
on presently is a quantum point contact (QPC) detector measuring a
double quantum dot charge qubit (DD), a system that has been
extensively investigated, both theoretically\cite{DDth} and
experimentally.\cite{DDexp}

A generic problem that arises if one is interested in making a
projective measurement, made out of many weak measurements, is
that the dynamics from the Hamiltonian evolution combines in a
nontrivial way with the measurement dynamics. The way around this
problem is with quantum nondemolition (QND)
measurements.\cite{book}  The QND scheme employed in this paper is
that of kicked QND
measurements\cite{ANJkick1,resonator,strobe1,strobe2,ANJkick2,averinflux}
on a qubit.  This idea was discussed for the QPC in
Ref.~\onlinecite{ANJkick1}, and was inspired by a similar rotating
QND scheme of Ref.~\onlinecite{averinqnd}. By turning the detector
on and off in a time scale much faster than the Rabi oscillation
period, the detector gives a little information about the quantum
state. The qubit is then allowed to make a full Rabi oscillation
before the next weak measurement, so the measured observable is
static in time from the perspective of the measurement device,
effectively turning off Hamiltonian evolution.  In this
fashion, the information contained in the qubit is teased out over
many measurements, or detector kicks.  Kicked QND measurements
have many advantages that recommend them as a technique of choice
both for theoretical treatment, as well as for experimental
implementation:
\begin{itemize}
\item{Theoretically, the kicked measurements may be described with
a non-unitary quantum map.  The map is discrete in the time index,
but the measurements are weak, not projective.  The quantum map
formalism allows for a technically simple treatment of the
combined Hamiltonian and measurement-induced dynamics in a global
manner, in contrast to continuous measurement analysis using
conditional differential Langevin equations.}
\item{The kicking mechanism is convenient, in that it allows the
measurement strengths to be fully tunable by adding more kicks, as
well as the easy inclusion of unitary operations by simply waiting
a fraction of a Rabi period.}
\item{Kicked QND measurements may be implemented in experiments
with several advantages.  The kicks may be accomplished with a
pulse generator on a QPC measuring a DD, and the waiting time
between kicks gives external decision circuitry the needed time to
process the data in order to do real-time feedback. Also, the pump
variation introduced in Ref.~\onlinecite{ANJkick2} removes the
uninteresting background signal of the measurement, and just gives
the bare output signal as either positive or negative pumped
current.}
\end{itemize}

The purpose of this paper is two-fold.   The first topic is
formal: To start with the well-known POVM approach to generalized
measurements, and derive the quantum Bayesian formalism from it,
starting with a scattering detector.  The detector physics allows
a natural translation of the abstract POVM formalism into physical
processes, and the quantum Bayesian formalism is recovered in the
weak coupling limit.  After discussing kicked-QND measurements,
we show how both kicked measurements and unitary operations may be
recast in terms of conformal maps, and demonstrate a close
parallel with the mathematics of the special theory of relativity.

The second topic is physical: the formal results are applied to
make predictions using conditional operations with real-time
feedback: (1) We derive an algorithm to deterministically produce
an arbitrary pure state after a (random) weak measurement using
feedback. (2) We investigate the purification process under
measurement and generalize Jacobs' qubit feedback protocol to
speed up purification with feedback.\cite{jacobs}

The paper is organized as follows.  In Sec. II, we derive the
quantum Bayesian formalism from POVMs applied to a mesoscopic
scattering detector in the weak coupling limit.  Kicked QND
measurements are reviewed in Sec. III, in the context of the
quantum Bayesian formalism.  In Sec. IV we introduce a
stereographic projection representation, and rewrite the
measurement dynamics as a stochastic conformal mapping.  A close
analogy to special relativity is also discussed.  Sec. V combines
kicked measurements with unitary operations, and calculates
measurement statistics.  Sec. VI introduces conditional phase
shifts in order to deterministically produce the same quantum
state after a measurement. In Sec. VII, we investigate the
purification process of any initially mixed density matrix under
kicked measurement. Sec. VIII contains our conclusions.

\section{Derivation of the quantum Bayesian formalism from POVM}
\label{POVMbayes}

The formalism used in this paper is called the quantum Bayesian
approach \cite{bayesian} because it may be considered as a
generalization of classical Bayesian inference.  An analogous
approach to quantum measurement that is better known in the
quantum information community has been given the unfortunate name
of positive operator-valued measure (POVM)
formalism.\cite{povm,nielsen} In this section, the quantum
Bayesian formalism for a solid state system is derived from POVMs.

Consider a bipartite system composed of $A$ and $B$, where the
states of $B$ are expressed in the orthonormal basis $\vert Q
\rangle_B$, and the states of $A$ are expressed in the orthonormal
basis $\vert j \rangle_A$.  A unitary transformation that
entangles the states in $A$ with the states in $B$ is given by \be
\vert \psi \ra_A \vert 0 \rangle_B \rightarrow \sum_{Q} {\bf M}_Q
\vert \psi \ra_A \vert Q \rangle_B, \label{unitary} \ee where we
consider an initial state $\vert \psi \ra_A$ in $A$, described by
a density operator ${\bf \rho}_A$, an initial state $\vert 0
\rangle_B$ in $B$, and have introduced the measurement operators
${\bf M}_Q$ that are indexed by the states in $B$, and operate in
$A$.  The normalization of the states gives the completeness
relation, $\sum_{Q} {\bf M}^\dagger_Q {\bf M}_Q={\bf 1}$.  Now
make a projective measurement on $B$ alone, and find the result
$Q$.  Any measurement of this kind may be described as a POVM in
A.  The probability of finding the result $Q$, called $P(Q)$, is
given by \be P(Q) = {\rm Tr}( {\bf \rho}_A {\bf M}^\dagger_Q {\bf
M}_Q), \label{pofq} \ee while the outcome of this measurement
prepares a new density operator of $A$, conditioned on the outcome
$Q$, and is given by \be {\bf \rho}'_A = \frac{ {\bf M}_Q\, {\bf
\rho}_A \, {\bf M}^\dagger_Q}{ {\rm
   Tr}({\bf \rho}_A {\bf M}^\dagger_Q {\bf M}_Q)}.
\label{update} \ee This defines a mapping ${\bf \rho}'_A = {\bf
\$}({\bf \rho}_A)$ from density operators to density operators,
known as a super-operator, which is not unitary in general.

While the above results are standard generalizations of projective
measurement on $A$, the abstract formulation obscures how to
practically apply the POVM to a specific physical system.  We now
consider such a system in the solid-state: the quantum point
contact (QPC), measuring the state of a double quantum dot (DD),
in order to see how the POVM translates into physical language.
The Coulomb interaction between the DD and QPC alters the
transport properties of the QPC, and can thus be used to detect
which quantum dot the DD electron occupies. The QPC is described
with the help of a scattering matrix $S_j$ that depends on the
physical state of the DD.  Following Averin and Sukhorukov,
\cite{AS} the unitary evolution of the total state during the
scattering process is comprised of the state of an individual
electron (system $B$) incident from the left electrode, $\vert
in\ra_B$, and the state of the DD (system $A$), $\alpha \vert 1
\ra_A + \beta \vert 2\ra_A$, evolving as
\bea &\vert in \ra_B (\alpha \vert 1 \ra_A + \beta \vert 2\ra_A)
\rightarrow & \label{cohevo}
\\ &\alpha (r_1 \vert L\ra_B + t_1 \vert R \ra_B) \vert
1\ra_A + \beta (r_2 \vert L\ra_B + t_2 \vert R \ra_B) \vert
2\ra_A,& \nonumber \eea
where $\vert L,R \ra_B$ represent scattering states of the QPC
that have either been reflected or transmitted, and the
transmission and reflection amplitudes $t_j$ and $r_j$ are
elements of the scattering matrix $S_j$: \be S_j  =
\begin{pmatrix} r_j & {\bar t}_j \\ t_j & {\bar r}_j
\end{pmatrix}.
\label{s} \ee The state of the ``logical'' qubit of the DD, is now
entangled with the ``ancilla'' qubit of the left/right position of
the QPC electron, and this comprises the bipartite system. Using
the evolved state (\ref{cohevo}), we can now read off the
measurement operators, ${\bf M}_Q$, of this unitary operation in
the $(\vert 0\ra_A, \vert 1\ra_A)$ basis, \be {\bf M}_{L} =
\begin{pmatrix} r_1 & 0 \\ 0 & r_2
\end{pmatrix}, \qquad
{\bf M}_{R} =  \begin{pmatrix} t_1 & 0 \\ 0 & t_2
\end{pmatrix},
\label{M} \ee and easily verify that ${\bf M}^\dagger_L {\bf M}_L
+ {\bf M}^\dagger_R{\bf M}_R= {\bf 1}$ from probability
conservation.  Counting the electron in the collector of the QPC
gives a random outcome, $Q=1$ if the electron is counted, or $Q=0$
if the electron is not counted, and makes a projective measurement
on the $B$ part of the Hilbert space.
%We thus identify the physical scattering state, with both logical and
%counted electron states, $\vert L \ra_B \rightarrow \vert 0\rangle_B$,
%and $\vert R \ra_B \rightarrow \vert 1\rangle_B$.
Equation (\ref{pofq}) gives the probability of counting the
electron (or not), \be P(1) = \rho_{11} T_1 + \rho_{22} T_2,
\qquad P(0) = \rho_{11} R_1 + \rho_{22} R_2, \label{count} \ee
where $\rho_{ij}$ are the elements of the DD density matrix in the
$\vert 1,2\ra$ basis, $T_{j} = \vert t_{j}\vert ^2$, $R_{j} =
\vert r_{j}\vert ^2$, and $T_j +R_j =1$.  The density matrix of
the DD qubit may be updated, given the outcome of the measurement
with Eq.~(\ref{update}).  If $Q=1$, so an electron is counted,
then
%\begin{widetext} \onecolumngrid
%
\bea \rho'_{11} &=& T_1 \rho_{11}/P(1),\quad \rho'_{22} =
1-\rho'_{11}, \nonumber \\ 
\rho'_{12} &=&  (\rho'_{21})^\ast = t_1
t_2^\ast \, \rho_{12}/P(1) \nonumber \\
&=& \rho_{12}\, e^{i\xi} \sqrt{\rho'_{11}
\rho'_{22}/\rho_{11} \rho_{22}}, \label{yes} \eea
where $\xi$=
Arg$(t_1 t_2^\ast)$; while if $Q=0$, so an electron is not
counted, or equivalently, a hole is counted, then
\bea \rho'_{11} &=&
R_1 \rho_{11}/P(0),\quad \rho'_{22}= 1-\rho'_{11}, \nonumber \\
\rho'_{12} &=&   (\rho'_{21})^\ast = r_1 r_2^\ast \, \rho_{12}/P(0) \nonumber \\
&=& \rho_{12}\, e^{i\chi} \sqrt{\rho'_{11} \rho'_{22}/\rho_{11}
\rho_{22}}, \label{no}
\eea where $\chi$= Arg$(r_1 r_2^\ast)$.
%\end{widetext}\twocolumngrid

The results (\ref{yes},\ref{no}) have a natural interpretation as
a quantum Bayes formula: The diagonal density matrix elements are
interpreted as classical probabilities, and are updated according
to the classical Bayes formula, while the off-diagonal elements
have a more exotic rule.\cite{bayesian} Note that if the initial
DD qubit state is pure, it remains pure after the measurement.
This is because while the entanglement enlarged the effective
Hilbert space which would lead to decoherence if the entangled
information went undetected, the measurement of the QPC electron
collapses the 2-particle state back down to a different pure DD
state.

It is instructive to contrast the POVM procedure with the well
known ``decoherence'' approach to quantum measurement in this most
simple case.  The decoherence approach corresponds to explicitly
averaging the elements of the density matrix over all possible
outcomes of the detector.  In this case, the two possible outcomes
of the measurement (\ref{yes},{\ref{no}) are used to obtain \be
\la \rho'_{11} \ra = P(0) \rho'_{11}(0) + P(1) \rho'_{11}(1) =
\rho_{11} \label{ond} \ee for the diagonal elements ($\rho_{22} =
1-\rho_{11}$), and \be \la \rho'_{12} \ra = P(0) \rho'_{12}(0) +
P(1) \rho'_{12}(1) = (t_1 t_2^\ast + r_1 r_2^\ast) \rho_{12}
\label{offd} \ee for the off-diagonal elements ($\rho_{21}
=\rho_{12}^\ast$), in agreement with Averin and
Sukhorukov.\cite{AS,note1} The new off-diagonal matrix elements
are reduced because $\vert t_1 t_2^\ast + r_1 r_2^\ast \vert \le
1$, resulting in effective decoherence, while the diagonal matrix
elements are preserved.  The predictive advantage of the quantum
Bayesian approach comes from not averaging over the measurement
results, but rather conditioning the quantum density matrix on the
result obtained in a particular physical realization.

The above POVM analysis is not difficult to extend to $M$
``ancilla'' qubits, or QPC electrons.  The basis $\vert Q \ra$ is
now spanned by $M$ qubits, each being projected to either $0$ or
$1$. Rather than find the probability of obtaining a given
sequence of $0$s and $1$s in the output, it happens that it is
sufficient to find the probability of just obtaining the total
charge $n=\sum_{i=1}^M Q_i$, given $M$ attempts, where $Q_i=(0,
1)$.  In other words, sequence does not matter, only the total
number of counted electrons. This mapping is illustrated in Fig.
1, where the ancilla outcome 1 or 0, is mapped respectively into
either counting an electron, or not counting an electron in the
current stream.  The generalization of the quantum Bayesian rules
(\ref{count}-\ref{no}) for M ancilla qubits is done by replacing
the success probabilities $T_j$, and the failure probabilities
$R_j=1-T_j$ by the probability $P(m, M\vert j)$ to measure $m$
electrons in $M$ attempts, under the condition that the qubit is
in state $\vert j\ra$,
\be P(m, M\vert j) = \begin{pmatrix} M \\
m\end{pmatrix} T_{j}^m (1-T_{j})^{M-m}, \label{binomial} \ee
which is the binomial distribution.

While this is sufficient for the diagonal matrix elements, the
off-diagonal matrix elements get a different phase shift after
each electron is measured.  The total phase shift is $\Phi(m, M) =
M\chi + m (\xi-\chi)$, which is deterministic if $\xi=\chi$.  The
symmetric quantum point contact has the property that
$\xi=\chi=0$,\cite{dephaseth,DDth} so this simplification will be made in
the rest of the paper.  For many electrons $M$, the current is
determined by $m/M$. The physically relevant weak-coupling (weakly
responding) limit corresponds to $(T_1- T_2)/(T_1+ T_2)\ll 1$.
Appealing to the central limit theorem, we treat the detector shot
noise in the Gaussian approximation with little lost
information.\cite{noteFCS} The QPC detector is efficient, in the
sense that no information about the DD qubit is lost in the noisy
current output.\cite{DDth,bayesian}
This is analogous to saying in the
logical language that no ancilla qubit was left unprojected, and
that the unitary operations did not hide any qubit information in
the phase of the ancilla qubits that is destroyed after projection
in the left/right basis.  
\begin{widetext} \onecolumngrid
\begin{figure}[t]
\label{fig1} \epsfxsize=11cm \center{\epsfbox{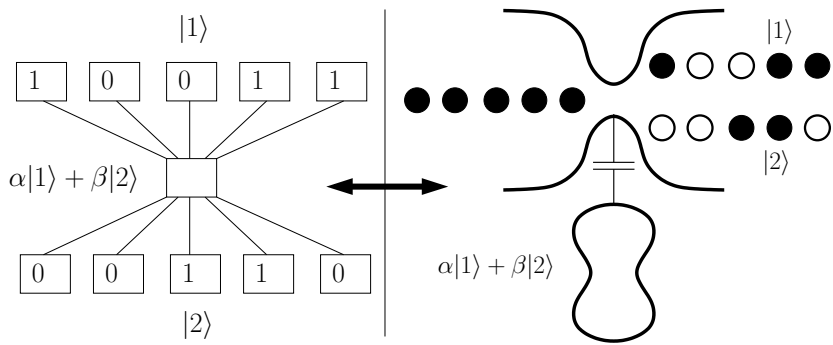}}
\caption{Weak entangling of a logical qubit with many ancilla
qubits, followed by projective measurement on the ancilla qubits
can be mapped onto the measurement of a quantum double dot by the
transport electrons of a quantum point contact.  Two given
measurement realizations are shown, depending on whether the
logical qubit state is either $\vert 1\ra$ or $\vert 0\ra$.  The
random ancilla result $1$ is mapped onto measuring an electron in
the current collector (denoted with a filled circle), while the
random ancilla result $0$ is mapped onto not measuring an electron
in the current collector, or equivalently, measuring a hole
(denoted with an empty circle).  The fact that the states $\vert
1\ra$ and $\vert 2\ra$ of the logical qubit alters the probability
of projecting the ancillas to 0 or 1, allows a weak POVM
measurement on the logical qubit.}
\end{figure}
\end{widetext}\twocolumngrid
One manifestation of an efficient
detector in the dephasing approach, is that the measurement rate
coincides with the measurement-induced dephasing
rate.\cite{DDth,bayesian} Ref.~\onlinecite{AS} related the
measurement rate to one of the R\'enyi entropies, or the
statistical overlap $O_{1,2} = \sum_m [P(m, M\vert 1) P(m, M\vert
2)]^{1/2}$, and showed that for the symmetric QPC it coincides with 
the reduction of the off-diagonal matrix elements. 
It is one feature of the quantum
Bayesian approach that this particular measure comes out in a
natural way. To see this, we first note a general property of
the density matrix, $\vert \rho_{12} \vert  \le
\sqrt{\rho_{11} \rho_{22}}$, that simply comes from the density
matrix eigenvalues being bounded between zero and one. Next, we
consider an initially pure state ($\vert \rho_{12} \vert  =
\sqrt{\rho_{11} \rho_{22}}$) and notice that the elements of the
density matrix after a measurement, $\rho'$, obey the relation
%
%\begin{widetext} \onecolumngrid
\be \left\vert \frac{\rho_{12}' }{ \rho_{12} }\right\vert  \le
\sqrt{\frac{\rho_{11}' \rho_{22}'}{\rho_{11} \rho_{22}}}.
\label{ratio} \ee  The above relation is valid for every given
measurement outcome, so it is also valid after averaging over the
distribution of results, $\la {\cal O}(m) \ra =\sum_m P(m,M) {\cal
O}(m)$, where ${\cal O}$ is any observable, and $P(m,M) =
\rho_{11} P(m, M\vert 1) + \rho_{22} P(m, M\vert 2)$. Using the
classical Bayes rule for the diagonal elements, taking ${\cal O}
=\vert \rho_{12}' / \rho_{12} \vert$, and the fact that $\vert \la
{\cal O} \ra \vert \le \la \vert {\cal O} \vert \ra$, we obtain
the generalized efficiency relation
\be \left \vert \left \la \frac{\rho_{12}' }{ \rho_{12} }\right
\ra \right \vert \le \sum_m \sqrt{ P(m, M\vert 1)  P(m, M\vert
2)}. \label{ger}\ee
Notice this relation is quite general, as no particular update 
rule for the off-diagonal matrix element has been invoked.
Therefore, a detector reaching the upper bound (\ref{ger}) can
naturally be called ideal, or 100\% efficient. (For the appropriate definition of
efficiency for an asymmetric detector see, e.g., Refs.~\onlinecite{bayesian,ank}.)

Thus far, we have focused only on the dynamics of the measurement
process, and have neglected the Hamiltonian evolution of the DD
qubit.   This evolution rotates the quantum state, and continually
changes the effective measurement basis, which typically ruins the
desired continuous measurement.   The way to get around this
bothersome detail is with QND measurements, the subject of the
next section.

\section{kicked QND measurements}
\label{kick}

The unifying theme behind all QND schemes is to couple the
measurement apparatus to the qubit with an operator that is an
approximate constant of motion of the measured quantum
system.\cite{book} In this way, the detector only measures the
state in the desired fixed basis, and the internal quantum
dynamics that would otherwise spoil the desired measurement is
circumvented. The specific scheme we employ in this paper is that
of kicked QND measurements, introduced by V. Braginsky {\it et
al.} \cite{strobe2} and K. Thorne {\it et al.} \cite{strobe1} for
the harmonic oscillator. In Ref.~\onlinecite{ANJkick1}, this idea
is introduced for two-state systems by making an analogy to a cat
playing with a string that moves in a circle.  In the kicked QND
mode, the cat sits in one spot waiting for the string to come to
it, and only then bats at it.  The motion in a circle comes from
the simple Hamiltonian evolution of a two-state system.  If $H=
\epsilon \sigma_z/2 + \Delta \sigma_x/2$ is the qubit Hamiltonian,
where $\Delta$ is the tunnel coupling energy, and $\epsilon$ is
the energy asymmetry, then unitary evolution for a time $t$ is
given by
\bea {\bf U} &=& \exp[-i t(\epsilon \sigma_z + \Delta \sigma_x)/2]
= {\bf 1} \cos(Et/2) \label{coherent} \\
&-& i \sigma_z (\epsilon/E) \sin(Et/2) - i \sigma_x (\Delta/E)
\sin(Et/2), \nonumber \eea
where $E=\sqrt{\epsilon^2 +\Delta^2}$, and $\hbar=1$ throughout
the paper. From the perspective of the qubit, the measurement
apparatus only measures at approximately discrete points in time.
In this reduced problem, by choosing the waiting time between
kicks to be $\tau_q = 2 \pi/E$ (or some integer multiple $n$
thereof), the unitary evolution (\ref{coherent}) becomes ${\bf U}
\rightarrow (-{\bf 1})^n$. The operator we want to measure is then
static in time, and is thus a QND measurement. (The evolution is
also simple if $n$ is a half-integer, especially if $\epsilon=0$).
However, from the point of view of the detector, the on/off pulse
lasts much longer than any detector time scale, so many electrons
pass through the QPC. If $\tau_0$ is the time scale of the QPC
electron correlation, $\tau_V$ the time scale of the pulse
duration, and $\tau_q$ is the Rabi oscillation period, then the
considered time scale ordering is $\tau_0 \ll \tau_V \ll
\tau_q$.\cite{higher}

\begin{figure}[t]
\begin{center}
\leavevmode \psfig{file=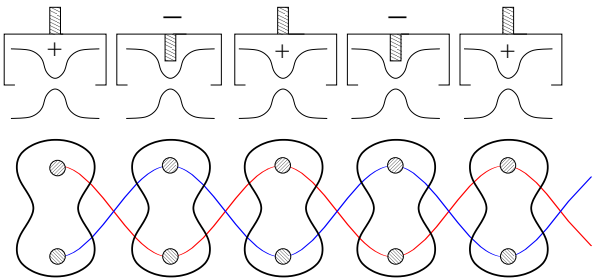,width=8cm} \caption{(color
online). (After Ref.~\onlinecite{ANJkick2}). Visualization of the
kicked QND measurement scheme. A voltage pulse is applied to the
quantum point contact on a time scale $\tau_V \ll \tau_q$,
followed by a quiet period of zero voltage bias, lasting for a
Rabi oscillation period $\tau_q$, followed by another pulse, and
so on. The up/down variation is depicted, where the kicks come
every half period, and the sign of the voltage pulse alternates
with every kick.  In this scheme, qubit read-out is by simply
measuring the sign of the current, and corresponds to an
elementary quantum pump.} \label{fig2}
\end{center}
\vspace{-5mm}
\end{figure}
A pump variation on kicked QND measurements was given by B\"uttiker and the
authors in Ref.~\onlinecite{ANJkick2}, where instead of giving the
same kick every Rabi oscillation, the experimentalist gives a
sequence of voltage kicks to the QPC with a pulse generator,
alternating in sign, every {\it half} oscillation period (see Fig.
2). In this scenario, we have shown that if $\epsilon=0$, qubit
readout is accomplished by pumping current: the kicks provide one
AC current source, and the dynamics of the qubit provides another
(intrinsically quantum mechanical) AC current source, that
nevertheless causes a net DC current flow in the
QPC.\cite{pump,bitflip} There are two limiting cases the system is
driven into:  Either the qubit oscillations are of the same phase
with the pump oscillations, pumping positive current, or the qubit
oscillations are of opposite phase with the qubit oscillations,
pumping negative current.

To characterize the result of each measurement kick,  the
parameters of the measurement process with an ideal QPC detector
are specified by the currents, $I_1$ and $I_2$, produced by the
detector when the qubit is in state $\vert 1\ra$ or $\vert 2\ra$,
and the detector shot noise power $S_I =e I (1-T)$ (where $T$ is
the transparency).\cite{noteconv} The typical integration time
needed to distinguish the qubit signal from the background noise
is the measurement time $T_M = 4S_I/(I_1-I_2)^2$.  Shifted,
dimensionless variables may be introduced by defining the current
origin at $I_0=(I_1+I_2)/2$, and scaling the current per pulse as
$I -I_0= x (I_1-I_2)/2$, so $I_{1,2}$ are mapped onto $x=\pm 1$.
The weak static coupling (per pulse) between QPC and DD implies
that the kick duration $\tau_V$ is less than the measurement time
$T_M$. We take $x$ to be normally distributed with variance $D =
T_M/\tau_V$.  The typical number of kicks needed to distinguish
the two states is $D$, where we assume $D\gg 1$.

The measurement result $\cal I$ after $N$ kicks is
\be {\cal I} = \frac{1}{N}\sum_{n=1}^{N} x_n, \ee
and we seek the conditional probability distribution $P({\cal I},
N \vert \rho)$ of measuring the dimensionless current ${\cal I}$,
starting with a given density operator $\rho$ prepared before the
first kick. The functions $P({\cal I},N|j)$ are defined as
classical probability distributions of the current with mean
${\cal I} =\pm 1$ (if $j=\pm 1$) and variance $\sigma^2=D/N$, the
Gaussian equivalent of (\ref{binomial}),
\bea P({\cal I},N|1) &\equiv& \frac{1}{\sqrt{2\pi D/N}}
\exp\left[- \frac{({\cal I} - 1)^2}{2D/N}\right], \nonumber \\
P({\cal I},N|2) &\equiv& \frac{1}{\sqrt{2\pi D/N}} \exp\left[-
\frac{({\cal I}+1)^2}{2D/N}\right] , \eea
and the notation $P_{j}(x_n)$ is adopted for the $j=1,2$
distributions of the $n$th kick. The probability density of
measuring the result $x_n$ after one kick is determined by the
state of the qubit just before the measurement, and is given by
the analog of (\ref{count}),
\be P(x_n) = \rho_{11}^{(n)} P_1(x_n) + \rho_{22}^{(n)} P_2(x_n).
\label{currentprob} \ee
The density matrix of the qubit is updated based on the
information obtained from the measurement that just occurred. This
is done with the quantum Bayesian update rules,\cite{bayesian}
that defines a non-unitary quantum map directly analogous to
Eqs.~(\ref{yes},\ref{no}),
%
%\begin{widetext} \onecolumngrid
\bea \rho_{11}^{(n+1)} &=& \frac{\rho_{11}^{(n)}
P_1(x_n)}{\rho_{11}^{(n)}
 P_1(x_n) + \rho_{22}^{(n)} P_2(x_n)},\nonumber \\
\rho_{12}^{(n+1)} &=& \rho_{12}^{(n)}\sqrt{\rho_{11}^{(n+1)}
 \rho_{22}^{(n+1)}/\rho_{11}^{(n)} \rho_{22}^{(n)}}, \nonumber \\
   \rho_{22}^{(n+1)} &=& 1-\rho_{11}^{(n+1)}, \quad \rho_{12}^{(n+1)} 
= (\rho_{21}^{(n+1)})^\ast.
\label{bayesrules} \eea
This quantum map is a probabilistic, non-unitary relative of the
unitary maps studied in kicked quantum chaos.\cite{qmap1,qmap2}
The advantage of QND measurement in the Bayesian approach is seen
by using Eqs.~(\ref{bayesrules}) to express the conditional
probability density $P(x_n)$ in terms of the result of the
preceding kick $x_{n-1}$.   It follows from
(\ref{currentprob},\ref{bayesrules}) that
%
%\begin{widetext}
\be P(x_{n}) = \frac{\rho_{11}^{(n-1)} P_1(x_{n-1}) P_1(x_{n}) +  
\rho_{22}^{(n-1)} P_2(x_{n-1}) P_2(x_{n})}{P(x_{n-1})}.
\label{recursive} \ee
%\end{widetext}
%
This recursive relation helps in the calculation of the
(unconditional) probability distribution $P({\cal I}, N \vert
\rho)$ of finding current ${\cal I}$, starting with the density
matrix $\rho$, after $N$ kicks, given by
\be P({\cal I}, N \vert \rho) = \int \prod_{n=1}^N dx_n  P(x_n) \;
\delta\left({\cal I}-\sum_{i=1}^N x_i/N\right). \label{totp} \ee
Each application of (\ref{recursive}) generates a denominator that
cancels the probability density immediately preceding it in
(\ref{totp}).  Making $N$ iterations of (\ref{recursive}) gives
\begin{widetext}
\be P({\cal I}, N \vert \rho) =\int  \prod_{n=1}^N dx_n
\left[\rho_{11} P_1(x_1) \ldots  P_1(x_N) + \rho_{22} P_2(x_1)
\ldots  P_2(x_N)\right] \delta({\cal I}-\sum_{i=1}^N x_i/N) =
\rho_{11} P({\cal I},N \vert 1) + \rho_{22} P({\cal I},N \vert 2),
\label{qndanswer} \ee
\end{widetext}
where $\rho_{11}, \rho_{22}$ are the diagonal matrix elements of
the original density matrix, and the $N$ Gaussians compose to form
one Gaussian with a variance $N$ times smaller. As $N$ is
increased, the two qubit states can be distinguished with greater
statistical confidence, and eventually the distributions limit to
delta-functions, giving either ${\cal I}=1$ with probability
$\rho_{11}$, or ${\cal I}=-1$ with probability $\rho_{22}$.  A
one-sigma confidence is obtained when $N=D$, as previously stated.

It is worthwhile to point out several features of the above QND
measurement. First, $N$ weak measurements simply compose to make
an $N$-times stronger measurement.  Second, the QND measurement
output only involves the diagonal density matrix elements.  It is
for this reason that the output of a quantum nondemolition
measurement is equivalent to noisy classical measurement, where
the detected ``classical probabilities'' are given by the diagonal
density matrix elements in the preferred measurement basis. In
spite of the classical nature of the detector output, the qubit
state prepared after the $N$ measurements can be deduced from the
outcome of the random variable ${\cal I}$.  To characterize the
post-measurement density matrix, we note another recursion
relation from (\ref{bayesrules}),
\bea \rho_{11}^{(n+1)}/\rho_{22}^{(n+1)} &=& [\rho_{11}^{(n)}/
\rho_{22}^{(n)}]  [P_1(x_n)/P_2(x_n)] \nonumber \\
&=& [\rho_{11}^{(n)}/ \rho_{22}^{(n)}] \exp(2 x_n/D).
\label{trick} \eea
This result may be composed $N$ times, and the update of the
off-diagonal matrix element (\ref{bayesrules}) follows from
(\ref{trick}).  Using the definition ${\cal I}=(1/N) \sum_{n=1}^N
x_i$, the measurement of a given random current ${\cal I}$
prepares a new density matrix of the DD,
\be \rho' = \frac{1}{ \rho_{11}\, e^\gamma+ \rho_{22}\,
e^{-\gamma}}
\begin{pmatrix}
\rho_{11}\, e^\gamma & \rho_{12} \\
\rho_{12}^\ast   & \rho_{22}\, e^{-\gamma}
\end{pmatrix},
\label{rhonew} \ee
where $\gamma = {\cal I} N/D$ is the rescaled measurement result,
named the {\it rapidity} of the measurement, for reasons given in
the Sec.~\ref{conformal}.
\begin{figure}[t]
\begin{center}
\leavevmode \psfig{file=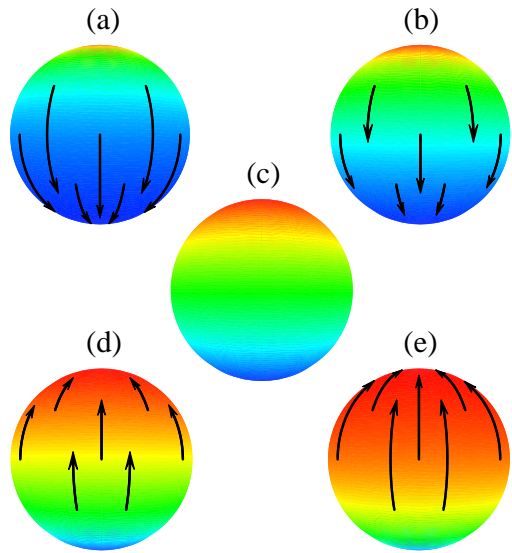,width=7.5cm}
\caption{(color online). (After Ref.~\onlinecite{ANJkick2}). The
conditional evolution of all initial pure states, represented on
the Bloch sphere, under continuous QND measurement by an efficient
detector. From (a-e), the {\it rapidity} of the measurement is
$\gamma= N {\cal I}/D =(-1, -.5, 0, .5, 1)$ respectively.  As the
detector obtains more information about the quantum state, we can
with greater statistical certainty distinguish the
post-measurement quantum state, so the Bloch sphere is more and
more red ($\vert 1\ra$) or blue ($\vert 2 \ra$), depending on the
value of the rapidity measured.  The conditional evolution of
several representative states is also indicated with black arrows.
The view-point is parallel to the equator of the Bloch sphere, so
$\vert 1 \ra \rightarrow$ North pole, and $\vert 2 \ra
\rightarrow$ South pole.} \label{fig3}
\end{center}
\vspace{-5mm}
\end{figure}
The conditional quantum dynamics of Eq.~(\ref{rhonew}) is
illustrated in Fig. 3, for all pure states, and the density matrix
is parameterized as $\rho = ({\bf 1} +\sum_i X_i \sigma_{i})/2$,
so $(X,Y,Z)$ give coordinates on the Bloch sphere. The $X$ and $Y$
behavior follows from $Z$, which is in turn conditioned on the
detector output ${\cal I}$, so the sphere is colored according to
the conditional evolution of $Z$,
\be Z' =\frac{\sinh \gamma + Z \cosh \gamma} { \cosh \gamma+ Z
\sinh \gamma}. \label{zrule} \ee
If the rapidity $\gamma$ is positive, then states are
``attracted'' toward the North pole, while if the rapidity
$\gamma$ is negative, then states are ``attracted'' toward the
South pole.  As the rapidity grows increasingly positive or
negative, we become more confident which state the qubit has
continuously collapsed to, but this also depends on the initial
state.  The conditional evolution of several representative states
is indicated with black arrows.

While the whole point of the kicked QND proposal was to
effectively turn off the qubit unitary evolution while the
continuous measurement is taking place, a much more interesting
situation arises when continuous (non-unitary) measurements are
combined with controlled unitary rotations.  The kicked
measurement set-up provides a simple way of generating a
single-qubit rotation: waiting.  Rather than spacing the pulses by
a full Rabi oscillation as previously described, we choose to wait
some fraction $r$ of a Rabi oscillation, $t_{\rm wait} = r
\tau_q$, that defines a phase shift $\phi = 2 \pi r$. A
single-qubit unitary operation (expressed in the z-eigenbasis),
\be {\bf U} =
\begin{pmatrix}
a & -b^\ast \\
b & a^\ast
\end{pmatrix},
\label{unitary2} \ee may be executed by choosing $r$, such that $a
= \cos(\phi/2) -i(\epsilon/E) \sin(\phi/2)$, and $b = -i(\Delta/E)
\sin(\phi/2)$, by using the Hamiltonian evolution of the qubit,
Eq.~(\ref{coherent}). Varying $r$, any point may be reached on a
circle on the Bloch sphere, which is fixed by $\epsilon$ and
$\Delta$. In order to reach any pure state (up to an overall
phase) by Hamiltonian evolution starting with a pure state, the
qubit asymmetry $\epsilon$ should also be varied between the kicks
with gate voltages.

\section{POVM measurement as a stochastic conformal map}
\label{conformal}

Before considering specific examples of weak measurement, combined
with unitary operations, we first reformulate the above set of
weak measurements (\ref{currentprob},\ref{bayesrules}), and
unitary operations (\ref{unitary2}) in an important special case:
where the initial state is pure, the detector is efficient (as
considered in this paper), so the post-measurement state is also
pure. The initial arbitrary DD state is defined as $\vert \psi \ra
= \alpha \vert 1 \ra+ \beta \vert 2 \ra$, with density matrix
elements $\rho_{11} =\vert \alpha \vert^2$, $\rho_{12} = \alpha
\beta^\ast$, $\rho_{21} = \rho_{12}^\ast$, $\rho_{22} = \vert
\beta \vert^2=1-\rho_{11}$.  Represented as coordinates on the
Bloch sphere, $(X,Y,Z)$, both measurements and unitary operations
leave the state on the surface, $X^2+Y^2+Z^2=1$.

Now make a stereographic projection of the Bloch sphere onto the
complex plane, with the complex variable $\zeta$, defined as
\be \zeta = \rho_{12}/\rho_{22} = (X+ i Y)/(1-Z) = \alpha/\beta.
\ee
The South pole of the Bloch sphere is identified as the origin of
the $\zeta$ plane, while the North pole is identified as $\infty$
on the $\zeta$ plane. Translating the unitary operation
(\ref{unitary2}) on the qubit into an operation on the complex
variable $\zeta$, we find the conformal mapping, \be \zeta' =
\frac{ a \zeta - b^\ast}{b \zeta + a^\ast}, \label{mobius} \ee
known as a M\"obius transformation.
%, that maps circles and lines to circles and
%lines. This last property is clear because a unitary operation
%rotates any circle on the Bloch sphere to another circle, which
%project to either circles or lines (great circles) on the complex
%plane. Thus, t
The group algebra of unitary rotations maps onto the group algebra
of conformal M\"obius transformations.  Reference
\onlinecite{qgeometry} points out that this property can be used
to demonstrate operator product identities for one qubit.

Translating the non-unitary Bayesian update equations for the
density matrix (\ref{bayesrules}), as an operation on the complex
coordinate $\zeta$, we find the following {\it stochastic
conformal mapping:} \be \zeta' =\zeta\, \sqrt{P_1(x)/P_2(x)} =
\zeta\, \exp( x /D). \label{scm} \ee This conformal mapping is
simply a random scale transformation with two fixed points: one at
$0$ (the South pole, state $\vert 2\ra$) and the other at $\infty$
(the North pole, state $\vert 1 \ra$). The random variable $x$ is
chosen from the probability distribution (\ref{currentprob}),
which translates to \be P(x) = \frac{\zeta P_1(x) +
(\zeta^\ast)^{-1} P_2(x)} {\zeta + (\zeta^\ast)^{-1}}.
\label{probconf} \ee Thus, any sequence of weak measurements,
combined with unitary operations can be translated into repeated
conformal mapping.
%, where the M\"obius mapping is
%deterministic, and the scaling is probabilistic.
We note that after an arbitrary sequence of weak measurements and
unitary operations, the definition $\zeta=\alpha/\beta$, together
with the normalization of the state, $\vert \alpha\vert^2
+\vert\beta\vert^2=1$, immediately allows the wavefunction to be
read off (up to an overall phase). The inclusion of asymmetric
measurements, where the phase shift in Eq.~(\ref{no}) is kept for
a broader class of scattering matrices may also be easily
included. This phase shift has the effect of twisting the Bloch
sphere proportionally to the value of current measured, which is
equivalent to including  phases in the scale factor, \be \zeta' =
\zeta\, \exp(i \theta_1) \exp[x(1+ i \theta_2)/D], \label{scm2}
\ee where $\theta_1,\theta_2$ correspond to the continuous limit
of the total acquired phase shift $\Phi(m,M) = M \chi + m
(\xi-\chi)$, described in Sec.~\ref{POVMbayes}.  The mapping
(\ref{scm2}) is obviously still conformal.

%\section{Relativity analogy}
%\label{rel}
If we momentarily let $x$ be a deterministic variable, then the
set of M\"obius mappings form a group, and the set of
deterministic scale transformations form a different group.  A
natural question that arises is what (if any) group is described
by the composition of (\ref{mobius}), with (\ref{scm})? Amusingly,
the answer is provided by the special theory of relativity.
Consider a relativistic four-vector $(X, Y, Z, T)$.  As is well
known, any element of the Lorentz group may be produced by making
a spatial rotation, followed by a boost in the (say) $Z$
direction, followed by another spatial rotation.  The boost from
$(X,Y,Z,T)$ to $(X',Y',Z',T')$ in the $Z$ direction may be
described as a hyperbolic rotation
\bea \begin{pmatrix} T' \\ Z' \end{pmatrix} &=& \begin{pmatrix}
\cosh \gamma & \sinh \gamma \\ \sinh \gamma & \cosh \gamma
\end{pmatrix} \begin{pmatrix} T \\ Z \end{pmatrix}, \quad
X'=X, \quad Y'=Y, \nonumber \\
\gamma &=& (1/2) \log\left(\frac{1+v}{1-v}\right), \label{boost}
\eea
where the rapidity $\gamma$ is introduced in terms of the velocity
parameter $v$, the physical velocity measured in units of the
speed of light, $c=1$.
%While $v \in (-1,1)$ has a finite domain, $\gamma \in
%(-\infty, \infty)$ has an infinite domain.

To connect this to spinor formalism, we follow the discussion in
Penrose and Rindler,\cite{penrose} and define a Hermitian
coordinate operator, \be {\bf C} =  \begin{pmatrix} T+Z & X+i\, Y
\\ X- i\,Y & T - Z \end{pmatrix}. \label{C} \ee Translating the
boost (\ref{boost}) into an operation on the coordinate operator
${\bf C}$ yields \be {\bf C'} ={\bf  A C A}^\dagger, \qquad {\bf
A} =
\begin{pmatrix} e^{\gamma/2} & 0 \\ 0 & e^{-\gamma/2} \end{pmatrix}.
\ee By fixing $T=1$, the celestial sphere $X^2+Y^2+Z^2=1$ is
defined. The celestial sphere is then stereographically projected,
defining the complex variable $\zeta = (X+ i Y)/(1-Z)$.  In the
complex plane, the boost is simply a scale transformation,
\be \zeta' = \zeta \, \exp \gamma. \label{boostscale} \ee
To extend the analysis to $N$ boosts with rapidities $\gamma_i$,
the mapping simply composes the $N$ scalings, to produce a boost
with rapidity $\gamma = \sum_{i=1}^N \gamma_i$. This conformal
mapping is identical with (\ref{scm}), the analogous quantum
measurement composition, if $\gamma=N {\cal I} /D$, the quantum
measurement parameter, is identified with the rapidity of the
boost. Furthermore, any spatial rotation of the sphere may be
interpreted as a unitary operation on the Bloch sphere, which
projects to the M\"obius mapping (up to an overall phase).  Thus,
the group described by the composition of (\ref{mobius}), with
(\ref{scm}) is the (restricted) Lorentz group.

The difference with the relativity analogy comes when we recall
that $\gamma=\sum_i x_i/D$ is a random variable.  The distribution
of this random variable (\ref{probconf}), explicitly depends on
the ``space-time'' coordinates $\zeta$, and thus breaks the
Lorentz invariance by introducing a preferred reference frame, the
$Z$-axis, with $Z=\pm 1$ as the attracting fixed points.  From the
quantum measurement point of view, this is a consequence of
choosing to measure along the $Z$-axis.   Therefore, for pure
states, the mapping (\ref{scm}, \ref{probconf}) may be viewed as a
stochastic Lorentz semi-group.

\section{Combined weak measurements and unitary operations}
\label{msm}

After having separately described weak measurement and unitary
operations, we now combine them. Consider an experiment, where
$N_1$ kicks are made, followed by a single qubit unitary operation
${\bf U}$ (produced by inserting a dislocation into the pulse
sequence), followed by $N_2$ kicks.  The measurement results
${\cal I}_1$ and ${\cal I}_2$ are defined as
\begin{equation}
{\cal I}_1=\frac{1}{N_1}\sum_{i=1}^{N_1} x_i, \qquad {\cal
I}_2=\frac{1}{N_2}\sum_{i=N_1+1}^{N_1+N_2} x_i.
\end{equation}
We seek the normalized probability distribution $P({\cal I}_1,
N_1; {\cal I}_2, N_2)$ of finding current ${\cal I}_1$ after $N_1$
kicks, and ${\cal I}_2$ after $N_2$ subsequent kicks.  This
distribution may also be interpreted as  joint counting
statistics.

The analysis from Sec.~\ref{kick} indicates that after the first
$N_1$ kicks, the measured current ${\cal I}_1$ will occur with a
probability given by (\ref{qndanswer}), and prepares a
post-measurement density matrix $\rho'$, given by (\ref{rhonew}),
described with the rapidity of the measurement $\gamma={\cal I}_1
N_1/D$.  The subsequent unitary operation ${\bf U}$
(\ref{unitary2}), (characterized by a phase $\phi$) rotates the
post-measurement density matrix,
\be
\rho^{\rm new} = {\bf U}\, \rho'\, {\bf U}^\dagger. \label{rot}
\ee
The following set of $N_2$ kicks start with the density matrix
(\ref{rot}), and continue to measure in the $z$-basis as before.
Equation (\ref{qndanswer}) may be applied again with the modified
initial density matrix (\ref{rot}) to deduce the (unconditional)
probability distribution of finding result ${\cal I}_1$ after
$N_1$ kicks, and result ${\cal I}_2$ after $N_2$ kicks,
%
%\begin{widetext} \onecolumngrid
\be P({\cal I}_1, N_1; {\cal I}_2, N_2) = P({\cal I}_1, N_1 \vert \rho)\times P({\cal I}_2, N_2 \vert \rho^{\rm new}),
%
%[\rho_{11} P({\cal
%I}_1,N_1 \vert 1) + \rho_{22} P({\cal I}_1,N_1)
% \vert 2)] \times  [\rho^{\rm new}_{11} P({\cal I}_2,N_2 \vert 1) +
%\rho^{\rm new}_{22} P({\cal I}_2,N_2 \vert 2)], 
\label{oneshift}
\ee where $P({\cal I}_i, N_i \vert \rho)$ is given in (\ref{qndanswer}), and
the new density matrix elements are given in terms of
the phase $\phi$ and rapidity $\gamma$ as (we let $\epsilon=0$ for
simplicity),
\begin{widetext}
\begin{eqnarray}
\rho^{\rm new}_{11} &=& \frac{ \cos^2(\phi/2) \rho_{11} e^\gamma +
\sin^2 (\phi/2) \rho_{22} e^{-\gamma}  -  2 \sin (\phi/2) \cos
(\phi/2)\; {\rm
 Im}\rho_{12}}{\rho_{11} e^\gamma + \rho_{22} e^{-\gamma}}, \qquad
\rho^{\rm new}_{22}= 1- \rho^{\rm new}_{11}, \nonumber \\
\rho^{\rm new}_{12} &=&  \frac{ {\rm Re} \rho_{12}+ (i/2)
\sin\phi\; (\rho_{11} e^\gamma -\rho_{22} e^{-\gamma}) +i \cos\phi
\; {\rm Im} \rho_{12}}{\rho_{11} e^\gamma + \rho_{22}
e^{-\gamma}}, \qquad  \rho^{\rm new}_{21}=(\rho^{\rm
new}_{12})^\ast, \label{rhonew2}
\end{eqnarray}
\end{widetext}
and the natural Hamiltonian dynamics performs the unitary
operation (\ref{coherent},\ref{unitary2}).  One interesting
feature of the result (\ref{oneshift},\ref{rhonew2}) is that the
outcome of the first $N_1$ measurements, ${\cal I}_1$, appears in
the expression involving the variables of the second set of kicks.
This immediately implies that the distribution does not factorize,
$P({\cal I}_1, N_1; {\cal I}_2, N_2) \ne P_1({\cal I}_1, N_1)
P_2({\cal I}_2, N_2)$.  The effect comes from the first set of
measurements preparing a given density matrix of the DD, which
affects the results of the next set of measurements.  From the
distribution (\ref{oneshift}), the average current in each
interval, as well as the correlation between the two may be
calculated:
%
%\begin{widetext} \onecolumngrid
\bea \la {\cal I}_1 \ra &=& \rho_{11} -\rho_{22}, \nonumber \\
\la {\cal I}_2 \ra &=& (\rho_{11} -\rho_{22}) \cos \phi -2 \exp(-N_1/2D) \sin
\phi\; {\rm Im}\, \rho_{12}, \nonumber \\ 
 \la {\cal I}_1 {\cal I}_2 \ra &=& \cos \phi. \eea
Also, as $N_1, N_2$ are taken to infinity in
(\ref{oneshift},\ref{rhonew2}), the distribution $P_{\rm PM}({\cal
I}_1, {\cal I}_2)$ from making simple projective measurements on
the DD is recovered,
\begin{eqnarray}
P_{\rm PM} &=& \rho_{11} \cos^2(\phi/2) \;
\delta({\cal I}_1 - 1)\delta({\cal I}_2 - 1) \nonumber \\
&+& \rho_{11} \sin^2(\phi/2) \; \delta({\cal I}_1 - 1)\delta({\cal I}_2 + 1) \nonumber \\
&+& \rho_{22} \sin^2(\phi/2) \; \delta({\cal I}_1 + 1)\delta({\cal
I}_2 - 1) \nonumber \\
&+& \rho_{22} \cos^2(\phi/2) \; \delta({\cal I}_1 + 1)\delta({\cal I}_2 + 1). 
\label{proj}
\end{eqnarray}
%\end{widetext}\twocolumngrid
%
It is now straightforward to generalize the result
(\ref{oneshift},\ref{rhonew2}) to any number of $m-1$ dislocations
in the pulse sequence, each of which has a phase shift of $\phi_k$
(and now $\epsilon$ is arbitrary),
\be P(\{{\cal I}_j,N_j\}) = \prod_{k=1}^m \left[\rho^{(k)}_{11}
P({\cal I}_k,N_k \vert 1) + \rho^{(k)}_{22} P({\cal I}_k,N_k \vert
2)\right] , \label{probgen} \ee
and each density matrix $\rho^{(k+1)}$ is defined in terms of the
density matrix $\rho^{(k)}$ after the previous dislocation,
\be \rho^{(k+1)} = {\bf U}_k \frac{1}{{\cal D}_k}
\begin{pmatrix}
\rho^{(k)}_{11} e^{\gamma_k} & \rho^{(k)}_{12} \\
\left[\rho^{(k)}_{12}\right]^\ast   & \rho^{(k)}_{22}
e^{-\gamma_k}
\end{pmatrix}
{\bf U}^\dagger_k, \label{manyrho} \ee
where ${\cal D}_k = \rho^{(k)}_{11} e^{\gamma_k}+ \rho^{(k)}_{22}
e^{-\gamma_k}$, the rapidities are $\gamma_k = {\cal I}_k N_k/D$,
the initial density matrix before the first kick is $\rho^{(1)}$,
and the matrix ${\bf U}_k$ has elements
%
%\begin{widetext} \onecolumngrid
%\be {\bf U}_k =
%\begin{pmatrix} \cos(\phi_k/2) - i (\epsilon/E) \sin(\phi_k/2) & - i
%(\Delta/E) \sin(\phi_k/2) \\ - i (\Delta/E) \sin(\phi_k/2) &
%\cos(\phi_k/2) + i (\epsilon/E) \sin(\phi_k/2)
%\end{pmatrix}.
%\label{manyU} \ee
%\end{widetext}\twocolumngrid
%
\bea ({\bf U}_k)_{11} &=& \cos(\phi_k/2) - i (\epsilon/E) \sin(\phi_k/2), \nonumber \\ 
({\bf U}_k)_{12}&=& - i (\Delta/E) \sin(\phi_k/2), \nonumber \\
({\bf U}_k)_{21}&=&  - i (\Delta/E) \sin(\phi_k/2), \nonumber \\
({\bf U}_k)_{22}&=& \cos(\phi_k/2) + i (\epsilon/E) \sin(\phi_k/2).
\label{manyU} \eea
Reference \onlinecite{ANJkick2} applied these results to violate a
Bell inequality in time.

\section{Conditional phase shifts and feedback protocols}
In the preceding Section, the phase shift $\phi$ was chosen
beforehand, independently of the result ${\cal I}_1$.  We can now
use the information gained in the first $N_1$ measurements, and
make a {\it conditional} phase shift, pending the outcome of the
random variable ${\cal I}_1$. This is essentially a feedback
protocol that the experimentalist can choose to execute, defined
by a function $\phi({\cal I}_1)$, so a different phase shift is
assigned to every possible random outcome of the continuous
measured current.

Kicked QND measurement provides a realistic mechanism for
implementing general qubit feedback protocols.  The reason for
this is two-fold: First, as seen in the previous section, any
combination of weak measurements and unitary operations may be
accomplished with a sequence of voltage pulses to the detector.
Second, the feedback circuitry must take the result obtained from
the measurement, execute logical operations, and command the
experimental apparatus to do something it otherwise would not have
done (like make a given phase shift). The intrinsic waiting time
between the kicks provides the needed time delay for all of the
above to take place.

We now explicitly find the feedback protocol $\phi({\cal I}_1)$ to
take a given pure state to any desired pure state after a weak
measurement. As the simplest case, consider any pure state on the
$Z-Y$ great circle of the Bloch sphere ($X=0,Y^2+Z^2=1$), and a
symmetric qubit, $\epsilon =0$.   Both Hamiltonian evolution, and
weak measurements (according to (\ref{rhonew})) do not take these
states out of the $Z-Y$ great circle. Therefore, knowing the
outcome ${\cal I}_1$, a conditional phase shift may be applied to
deterministically prepare {\it any} quantum state on the $Z-Y$
great circle of the Bloch sphere.  For definiteness, we choose to
shift to the state $\vert 1\ra$.  This choice has the advantage
that after the $N_2 \gg D$ measurements, the current will be
${\cal I}_2=1$ deterministically. The parametrization $(Y,Z) = (-
\sin \theta, \cos \theta)$ of the initial state is chosen, so that
if no measurement is made, the shift to the North pole may be done
with a phase shift $\phi =\theta$. The result (\ref{rhonew2}) is
applied by setting $\rho_{11}^{\rm new} =1$, and solving for
$\phi$ as a function of the rapidity $\gamma$, to find
\be \tan (\phi/2) = \tan (\theta/2)\; \exp(-\gamma).
\label{condshift} \ee
This answer interpolates between two extreme
strategies:\cite{note3} (1) If no measurement is made, just make
the desired phase shift, $\phi =\theta$. (2) If a projective
measurement is made, either do nothing if ${\cal I}_1=1$, or flip
the state by applying the phase shift $\phi = \pi$ if ${\cal
I}_1=-1$.  The asymptotic limits in the later case may be obtained
by expanding the inverse tangent to obtain
%
%\begin{widetext} \onecolumngrid
\be \phi/2 \approx \begin{cases} \tan \left( \theta/2\right)
\exp(-\gamma) & {\rm if}\; \gamma \gg 1,
{\rm and}\; \theta \ne \pm \pi, \\
\frac{\pi}{2}\; {\rm sign}(\theta) - \frac{\exp(\gamma)}{\tan (\theta/2)} & {\rm
if}\; \gamma \ll -1, {\rm and}\; \theta \ne 0.
\end{cases}
\ee
%\end{widetext}\twocolumngrid
%

The above real-time feedback proposal is experimentally promising
in the kicked scheme.  However, it demands fast time resolution
and feedback circuitry. An experimentally simpler proposal to
verify the above protocol is to make many realizations of weak
measurement, phase shift, weak measurement, where the phase shift
is chosen randomly in each realization. After the run is finished,
the data record may be reviewed, and all instances of phase shifts
where condition (\ref{condshift}) is approximately satisfied are
post-selected.  In this data subset, the prediction is that the
following set of $N_2\gg D$ kicks will deterministically find
${\cal I}_2=1$.

\section{Purification of initially mixed density matrices by weak measurement}
\label{pure} Under repeated weak QND measurements, eventually all
states collapse to either $\vert 1\ra$ or $\vert 2 \ra$, including
mixed initial states. However, the states $\vert 1\ra$ and $\vert
2 \ra$ are both pure, and therefore if the initial state is mixed,
a purification occurs during the measurement
process.\cite{korotkovpure}  This phenomenon is especially
counter-intuitive from the point of view of the dephasing approach
to quantum measurement. Jacobs has shown that the average
purification in a given time can be increased by the use of
continuous feedback.\cite{jacobs}  Jacobs' protocol is somewhat
counterintuitive for the qubit: always use Hamiltonian evolution
to rotate the state to $Z=0$, {\it i.e.} perpendicular to the
measurement axis.  The purpose of this section is (1) To show how
this idea can be easily implemented for our set-up, (2) To
demonstrate that the ``equatorial plane" protocol ({\it i.e.}
$Z=0$) is also optimal for kicked QND measurements which have a
continuous output of tunable measurement strength, and (3) To
generalize Jacobs' no-feedback purification solution to any
initial density matrix.

It is well known that any unitary operation preserves the purity
(or entropy) of the state.  It is interesting to note from
(\ref{bayesrules}), that during weak measurement there is a
different preserved physical quantity, that we name the ${\it
murity}$. For the qubit, the purity ${\cal P}$ and the murity $M$
are defined as
\be {\cal P}=X^2+Y^2+Z^2, \quad M = (X^2+Y^2)/(1-Z^2).
\label{defs} \ee
If the purity ${\cal P}=1$, then the murity $M=1$, reflecting the
statement made in Sec.~\ref{POVMbayes} that if the initial state
is pure, the post-measurement state is also pure.  We note that
${\cal P}$ may be expressed in terms of $M$ and $Z$ by ${\cal P}=M
(1-Z^2)+Z^2$.  After one kick, the change in the purity, $\Delta
{\cal P}$, (or purification) is given by
\be \Delta {\cal P} = {\cal P}'-{\cal P} = (1-M)[(Z')^2-Z^2], \ee
where we have used the fact that murity does not change during
measurement. Application of the quantum Bayesian update rules
(\ref{bayesrules}) yields $Z' = [\rho_{11} P_1(x) - \rho_{22}
P_2(x)] /(\rho_{11} P_1(x) + \rho_{22} P_2(x))$ (also given in
(\ref{zrule})), where $x$ is the measurement result, so the
purification is
\be \Delta {\cal P} = (1-{\cal P})\left(1 - \frac{1}{[\cosh (x/D)
+ Z\sinh (x/D)]^2}\right). \label{purification} \ee
Several observations are in order: First, if ${\cal P}=1$, the
purification $\Delta {\cal P}$ is automatically $0$, while the
first (deterministic) factor is maximal if ${\cal P}=0$.  Second,
if $x = 0$, the second (random) factor is zero, so there is no
purification, which corresponds to no gained information. Finally,
the first factor is between $[0,1]$, while the second factor could
be negative or positive, implying that either purification or
further mixing is possible in a given run.

The {\it average} purification is given by averaging
(\ref{purification}) over the distribution of $x$, given in
Eq.~(\ref{currentprob}), to yield
\be \la \Delta {\cal P} \ra = (1-{\cal P})[1 - f(D, Z)],
\label{avepure} \ee
where
%
%\begin{widetext} \onecolumngrid
\be f(D, Z) = e^{-\frac{1}{2D}} \int_{-\infty}^\infty \frac{dx}{\sqrt{2
\pi D}}\frac{ \exp(-x^2/2D)}{\cosh(x/D) + Z \sinh(x/D)}. \label{f}
\ee
%\end{widetext}\twocolumngrid
%
It is straightforward to check that $0\le f \le 1$, so there is
nonnegative average purification for all density
matrices.\cite{notef}  Changing variables to $\gamma=x/D$, it is
also straightforward to check asymptotic limits. Taking
$D\rightarrow 0$, the projective limit, $f(0, Z) = 0$ is
recovered, so $\la \Delta {\cal P}\ra =1-{\cal P}$, implying that
the final state is pure with unit probability. The opposite limit,
$D\rightarrow \infty$, corresponds to an vanishingly weak
measurement, so $f(\infty, Z) = 1$, or $\la \delta {\cal P}\ra
=0$, giving no purification.

The results (\ref{avepure},\ref{f}) allow us to find the optimum
average purification strategy for one kick.  The best strategy on
average is to rotate the density matrix to where the purification
(\ref{avepure}) is maximum.  These point(s) may be found by
maximizing $\la \Delta {\cal P} \ra$ on the Bloch ball, under the
constraint that $D$ and ${\cal P}$ are fixed.  The ${\cal P}$
constraint simply reflects the fact that unitary operations do not
alter the purity. This problem is equivalent to minimizing $f$
with respect to $Z$, by solving $df/dZ=0$, which leads to the
equation
\be  -\int_{-\infty}^\infty d\gamma \frac{ \exp(-D \gamma^2/2)
\sinh \gamma}{(\cosh \gamma + Z \sinh \gamma)^2} = 0. \ee
The solution is immediate because at the point $Z=0$, the
integrand is odd, so the integral is zero.  The fact that $f$ is
minimized at $Z=0$ is seen by noting that the integrand of
$d^2f/dZ^2(Z=0)$ is nonnegative. Therefore, the average
purification is maximized by applying a phase shift after the
first measurement that rotates the qubit to the equatorial plane
of the Bloch ball before the measurement. This result is in
agreement with Jacobs, who considered purification from a
two-outcome POVM of variable strength (and also the stochastic
Schr\"odinger equation limit).\cite{jacobs} Our approach is from
the complimentary perspective of a continuous outcome measurement of
variable strength.

Results (\ref{avepure},\ref{f}) have a simpler form in the large
$D$ limit, for very weak measurements. The average purification
$\la \Delta {\cal P} \ra$ and the noise in the purification, $\la
(\Delta {\cal P})^2 \ra$, are given to leading order in $D^{-1}$
as
\bea &\la \Delta {\cal P} \ra = (1-{\cal P})(1-Z^2)/D,& \nonumber
\\ &\la (\Delta {\cal P})^2 \ra = 4 (1-{\cal P})^2 Z^2/D.& \label{gp}
\eea
In order to compare purification with and without feedback,  we
first consider Jacobs' feedback protocol $Z=0$ at every time step.
Equation (\ref{gp}) implies that for this feedback protocol the
purification noise vanishes.  Therefore, the dynamical
purification is described by the deterministic rate equation
$d{\cal P}/dN = (1-{\cal P})/D$. Solving the equation with initial
condition ${\cal P}_0$ yields \cite{jacobs}
\be \la {\cal P}_N \ra = 1 + ({\cal P}_0-1) \exp(-N/D),
\label{feedback} \ee
showing an exponential approach to a pure state, with rate
$D^{-1}$.

In the no feedback case, the purity after $N$ kicks may be found
from the murity relation ${\cal P}' = M [1-(Z')^2] + (Z')^2$,
where $Z'$ is given in (\ref{zrule}). Averaging this relation over
the distribution (\ref{qndanswer}) yields the average purity after
$N$ kicks,\cite{notepur}
\begin{widetext} 
%\onecolumngrid
\be \la {\cal P}_N \ra = \sqrt{\frac{D}{2\pi N}}
\exp\left(-\frac{N}{2 D}\right) \int_{-\infty}^\infty d \gamma \exp\left(-\frac{D \gamma^2}{2 N}\right)
\frac{M (\cosh \gamma + Z
\sinh \gamma)^2 - (M-1) (\sinh \gamma + Z \cosh \gamma)^2}{\cosh
\gamma + Z \sinh \gamma}. \label{avepure2} \ee
\end{widetext}
After some manipulation, the above integral expression may
be simplified to
\bea \la {\cal P}_N \ra &=& 1 - (M-1)(Z^2-1)\sqrt{\frac{D}{2\pi N}}
\exp(-N/2D) \nonumber \\
&\times& \int_{-\infty}^\infty d\gamma \frac{\exp(-D
\gamma^2/2N)}{\cosh \gamma + Z \sinh \gamma}. \label{pursimp} \eea
%\end{widetext}\twocolumngrid
%
For large $N/D$, the dominant dependence comes from the term
outside the integral, and the $N/D$ dependence inside the
integrand may be neglected.  In this limit, the purity may be
approximated as
\be \la {\cal P}_N \ra \approx 1 - \pi (1-M)\sqrt{1-Z^2}
\sqrt{D/(2\pi N)} \exp(-N/2D), \label{limitpur} \ee
yielding an approach to purity with rate $(2D)^{-1}$, half as fast
as the feedback case, in agreement with Jacobs.  The result
(\ref{pursimp}) generalizes Jacobs' no-feedback purification
result to arbitrary initial states. Before concluding, we point
out that Wiseman and Ralph have recently shown that the advantage
of feedback for purification depends on how the question is
formulated.\cite{wiseman} If instead of asking about the average
purification for a fixed time, we ask about the average time taken
to reach a given purity, then the no feedback case is actually
better.

\section{Conclusions}
\label{conc} The quantum Bayesian approach to the problem of
quantum measurement has been derived from POVM formalism, applied
to a mesoscopic scattering detector.   By considering an
elementary scattering event, measurement operators associated with
the successful or failed detection of the electron in the current
collector can be identified. We recover the quantum Bayesian
formalism in the continuous current approximation.

Kicked QND measurements have been analyzed within the quantum
Bayesian formalism.  We derive a quantum map representation that,
while discrete in the time index, describes a sequence of weak
measurements.  Unitary operations (easily implemented by waiting a
fraction of a Rabi period), together with kicked measurements, can
be represented as a sequence of conformal mappings, where the
unitary maps are deterministic, and the kicked measurement maps
are stochastic.  A close analogy exists between these quantum maps
and the Lorentz transformations of special relativity.

We have calculated the measurement statistics associated with
combined weak measurements, and unitary operations.  
These results are applied to find the feedback
protocol that deterministically takes a given pure state to any
other desired pure state after a weak measurement, using
conditional phase shifts.

Next, we have investigated the process of purification of mixed
density matrices under kicked QND measurement. The concept of
``murity" (the physical quantity that is preserved under
measurement) has been introduced, and applied to calculate the
change in the state's purity associated with a measurement.
Purification with and without feedback has also been investigated.

We stress that kicked QND measurements provides an experimentally
viable way of implementing ideas in quantum feedback: Any
combination of weak measurements and unitary operations can be
accomplished by applying a sequence of voltage pulses to the
detector, and the intrinsic quiet time between kicks allows the
necessary processing time for feedback to occur.

\section{Acknowledgments}
We thank Markus B\"uttiker for discussions, who collaborated in
the initial stages of this project.  This work was supported by
MaNEP, the Swiss National Science foundation, AFRL Grant No.
F30602-01-1-0594, AFOSR Grant No. FA9550-04-1-0206, and TITF Grant
No. 2001-055. (A.N.J.) and the NSA/ARDA under ARO grant
W911NF-04-1-0204 (A.N.K.).


\begin{thebibliography}{04}
\bibitem{dephaseexp}
E. Buks, R. Schuster, M. Heiblum, D. Mahalu, and V. Umansky,
Nature {\bf 391}, 871 (1998); D. Sprinzak, E. Buks, M. Heiblum,
and H. Shtrikman, Phys. Rev. Lett. {\bf 84}, 005820 (2000).

\bibitem{dephaseth}
S. A. Gurvitz, Phys. Rev. B {\bf 56}, 15215 (1997);
I. L. Aleiner, N. S. Wingreen, and Y. Meir, Phys. Rev. Lett. {\bf 79}, 3740 (1997);
Y. Levinson, Europhys. Lett. {\bf 39}, 299 (1997);
M. H. Pedersen, S. A. van Langen, and M. B\"uttiker, Phys. Rev. B {\bf 57}, 1838 (1998).

\bibitem{thother}
M. B\"uttiker and A. M. Martin, Phys. Rev. B {\bf 61}, 2737 (2000); Phys. Rev. Lett. {\bf 84}, 3386 (2000);
G. Seelig, S. Pilgram, A. N. Jordan, and M. B\"uttiker, Phys. Rev. B {\bf 68}, 161310(R) (2003).

\bibitem{DDth}
A. N. Korotkov and D. V. Averin, Phys. Rev. B {\bf 64}, 165310
(2001); H. S. Goan and G. J. Milburn, Phys. Rev. B {\bf 64},
235307 (2001); S. Pilgram and M. B\"uttiker, Phys. Rev. Lett. {\bf
89}, 200401 (2002); A. A. Clerk, S. M. Girvin, and A.~D. Stone,
Phys. Rev. B {\bf 67}, 165324 (2003); D. V. Averin, in ``Exploring
the Quantum-Classical Frontier: Recent Advances in Macroscopic and
Mesoscopic Quantum Phenomena'', Eds. J. R. Friedman and S. Han
(Nova Science, Huntington, NY, 2003); cond-mat/0004364; A.
Shnirman, D. Mozyrsky, and I. Martin, Europhys. Lett. {\bf 67},
840 (2004); A. A. Clerk and A. D. Stone, Phys. Rev. B {\bf 69},
245303 (2004); A. N. Jordan and M. B\"uttiker, Phys. Rev. Lett.
{\bf 95}, 220401 (2005); N. P. Oxtoby, P. Warszawski, H. M.
Wiseman, H. B. Sun, R.E.S. Polkinghorne, Phys Rev B {\bf 71},
165317 (2005).

\bibitem{DDexp}
T. Hayashi, T. Fujisawa, H.-Du Cheong, Y.-Ha Jeong, and Y. Hirayama,
Phys. Rev. Lett. {\bf 91}, 226804 (2003); J.R. Petta, A.C.
Johnson, C.M. Marcus, M.P. Hanson, and A.C. Gossard, Phys. Rev.
Lett. {\bf 93}, 186802 (2004); J.M. Elzerman, R. Hanson, L.H.
Willems van Beveren, B. Witkamp, L.M.K. Vandersypen, and L.P.
Kouwenhoven, Nature (London) {\bf 430}, 431 (2004); J.R. Petta,
A.C. Johnson, J.M. Taylor, E.A. Laird, A. Yacoby, M.D. Lukin, C.M.
Marcus, M. P. Hanson, and A.C. Gossard, Science {\bf 309}, 2180
(2005); F.H.L. Koppens, J.A. Folk, J.M. Elzerman, R. Hanson, L.H.
Willems van Beveren, I.T. Vink, H.P. Tranitz, W. Wegscheider, L.P.
Kouwenhoven, and L.M.K. Vandersypen, {\it ibid.} {\bf 309}, 1346
(2005); J. M. Elzerman, R. Hanson, J. S. Greidanus, L. H. Willems
van Beveren, S. De Franceschi, L. M. K. Vandersypen, S. Tarucha,
and L. P. Kouwenhoven Phys. Rev. B {\bf 67}, 161308(R) (2003); A.
K. H\"uttel, S. Ludwig, K. Eberl, J. P. Kotthaus, Phys. Rev. B
{\bf 72}, R081310 (2005); J. Gorman, E. G. Emiroglu, D.G. Hasko, and D.A.
Williams, Phys. Rev. Lett. {\bf 95}, 090502 (2005).

\bibitem{book}
V. B. Braginsky and F. Ya. Khalili, {\it Quantum Measurement}
(Cambridge University Press, Cambridge, U.K., 1992).

\bibitem{strobe2}
V.~B. Braginsky, Yu.~I. Vorontsov, and F.~Ya. Khalili, JETP Lett.
{\bf 27}, 276 (1978).

\bibitem{strobe1}
K.~S. Thorne, R.~W.~P. Drever, C.~M. Caves, M. Zimmermann, and
V.~D. Sandberg, Phys. Rev. Lett. {\bf 40}, 667 (1978).

\bibitem{ANJkick1}
A. N. Jordan and M. B\"uttiker, Phys. Rev. B {\bf 71}, 125333
(2005).

\bibitem{resonator}
R. Ruskov, K. Schwab, and A.~N. Korotkov,  Phys. Rev. B {\bf 71},
235407 (2005).

\bibitem{ANJkick2}
A. N. Jordan, A. N. Korotkov, and M. B\"uttiker, cond-mat/0510782.

\bibitem{averinflux}
D. V. Averin, K. Rabenstein, and V. K. Semenov, cond-mat/0510771.

%averin QND
\bibitem{averinqnd}
D.~V. Averin, Phys. Rev. Lett. {\bf 88}, 207901 (2002).

\bibitem{jacobs}
K. Jacobs, Phys. Rev. A {\bf 67}, 030301(R) (2003);  K. Jacobs,
Proc. of SPIE {\bf 5468}, 355 (2004); J. Combes and K. Jacobs,
Phys. Rev. Lett. {\bf 96} 010504 (2006).

\bibitem{bayesian}
A. N. Korotkov, Phys. Rev. B {\bf 60}, 5737 (1999); Phys. Rev. B
{\bf 63} 115403 (2001).

\bibitem{povm}
E. B. Davies, {\it Quantum Theory of Open Systems} (Academic,
London, 1976); K. Kraus {\it States, Effects, and Operations:
Fundamental Notions of Quantum Theory} (Springer, Berlin, 1983);
A. S. Holevo, {\it Statistical Structure of Quantum theory}
(Springer, 2001).

\bibitem{nielsen}
M.~A. Nielsen, and I.~L. Chuang, {\it Quantum Computation and
Quantum Information}, (Cambridge University Press, 2000).

\bibitem{AS}
D. V. Averin and E. V. Sukhorukov,  Phys. Rev. Lett. {\bf 95},
126803 (2005).

\bibitem{note1}
Equations (\ref{ond},\ref{offd}) may be obtained directly by
tracing out the QPC electron states.

\bibitem{noteFCS}
The POVM generalization of Ref.~\onlinecite{AS} considering
electron counting statistics and its interplay with quantum
information is left as a future project.

\bibitem{ank}
A. N. Korotkov, Phys. Rev. B {\bf 67}, 235408 (2003).

\bibitem{higher}
A further condition needed is that the kick should not excite the
DD electron into higher states. For two tunnel coupled systems,
like the DD, the level splitting between the lowest two energy
states (which determines the effective qubit Hilbert space) is
exponentially smaller than the typical level spacing of a single
dot, so this condition is not very restrictive.

\bibitem{pump}
P. W. Brouwer, Phys. Rev. B {\bf 58}, R10135 (1998); J. E. Avron, 
A. Elgart, G. M. Graf, and L. Sadun, Phys. Rev. B {\bf 62}, R10618 (2000);  
M. Moskalets and M. B\"uttiker, Phys. Rev. B {\bf 70}, 245305 (2004).

\bibitem{bitflip}
On a long time scale, there will be random bit-flip errors if the
kicks are imperfect delta-functions.\cite{ANJkick1} The bit-flip
errors eventually randomize the average current to zero.  This
situation may be rectified with a feedback loop,\cite{feedback}
where when the bit-flip error is detected, a ``dislocation'' is
introduced into the pulse sequence by kicking twice ``up" or
``down".  The inclusion of this feedback loop produces a
sustainable quantum pump for an arbitrarily long time. In this
paper, only ideal delta-function kicks on the qubit are
considered.

%feedback loop
\bibitem{feedback}
R. Ruskov and A. N. Korotkov, Phys. Rev. B {\bf 66}, 041401(R)
(2002).

%\bibitem{rmp}
%Y. Makhlin, G. Sch\"on, and A. Shnirman,
%Rev. Mod. Phys. {\bf 73}, 357 (2001).

\bibitem{noteconv}
Our convention for the spectral density is $\la I^2\ra -\la I
\ra^2 = \int_{-\infty}^{\infty} S_I(\omega) d\omega/(2\pi)$.

\bibitem{qmap1}
G. Casati, B.~V. Chirikov, J. Ford, and F.~M. Izrailev, {\em
Stochastic Behavior in Classical and Quantum Hamiltonian Systems}
Vol.~93 of {\em Lecture Notes in Physics} edited by G. Casati and
J. Ford (Springer-Verlag: Berlin) p 334 (1979).

\bibitem{qmap2}
S. Fishman, D.~R. Grempel, and R.~E. Prange, Phys. Rev. Lett. {\bf
49}, 509 (1982).

\bibitem{qgeometry}
J. Lee, C. H. Kim, E. K. Lee, J. Kim, and S. Lee, Quantum
Information Processing {\bf 1}, 129 (2002); quant-ph/0201014
(2002).

\bibitem{penrose}
R. Penrose and W. Rindler, {\it Spinors and space-time, Vol. 1}
(Cambridge University Press, 1984).

\bibitem{note3}
It is interesting to note that Eq.~(\ref{condshift}) also has a
direct relativistic analog: It coincides with the aberration
formula for incoming light rays at polar angle $\phi$ perceived by
an observer with rapidity $\gamma$ in the $Z$ direction, compared
to incoming light rays at polar angle $\theta$ to a stationary
observer.  See Ref.~\onlinecite{penrose}.

%purification results
\bibitem{korotkovpure}
A. N. Korotkov, Physica B {\bf 280}, 412 (2000).

\bibitem{notef}
The quantity $f$ may be expressed more generally as $f =
\int_{-\infty}^\infty d x [P_1(x) P_2(x)]/[\rho_{11} P_1(x) +
\rho_{22} P_2(x)]$.

\bibitem{notepur}
This purification result may be expressed more generally as $\la
{\cal P}_N \ra  = \int d {\cal I}\{ [\rho_{11} P_1({\cal I}) -
\rho_{22} P_2({\cal I}) ]^2 + 4 \vert \rho_{12} \vert^2 P_1({\cal
I}) P_2({\cal I})  \} /[\rho_{11} P_1({\cal I}) - \rho_{22}
P_2({\cal I})]$.

\bibitem{wiseman}
H. M. Wiseman and J. F. Ralph, quant-ph/0603062.

\end{thebibliography}
\end{document}